\documentclass[12pt]{article}
\begin{document}

\author{M. EL BAZ\thanks{{\protect\large moreagl@yahoo.co.uk}}, A. EL HASSOUNI%
\thanks{{\protect\large lhassoun@fsr.ac.ma}}, Y. HASSOUNI\thanks{%
{\protect\large y-hassou@fsr.ac.ma}} and E. H. ZAKKARI\thanks{%
{\protect\large hzakkari@hotmail.com}}. \and \vspace{1 cm} \\
Laboratory of theoretical physics, P.O.BOX 1014,\\
University Mohammed V, Rabat Morocco.}
\title{The two-parameter higher order differential calculus and curvature on a
quantum plane. \\
\vspace{1 cm}}
\date{}
\maketitle

\vspace{2cm}

\begin{abstract}
We construct an associative differential algebra on a two-parameter quantum
plane associated with a nilpotent endomorphism $d$ in the two cases $d^{2}=0$
and $d^3=0$ $(d^2\neq 0).$ The correspondent curvature is derived and the
related non commutative gauge field theory is introduced.
\end{abstract}

\vspace{3cm}


\vspace{0.5cm}


\newpage\

\section{Introduction}

Many authors \cite{kobayashi, wess1, berezinski, schirrmacher, coquereaux1,
coquereaux2} have studied differential calculus with nilpotency $d^2=0$ on
quantum spaces with one or two-parameter quantum group symmetry\cite{manin1,
manin2}.

Recently, there have been several attempts \cite{kerner1, kerner2, kerner3,
dubois, dubois1, dubois2} to generalize the classical exterior differential
calculus to the one with exterior differential satisfying $d^N=0,$ $N\geq 3.$
Such generalization is called $q$-differential algebra \cite{dubois1}. The
latter is an associative unital $N$-graded algebra endowed with a linear
endomorphism $d$ satisfying $d^N=0$ and the $q$-Leibniz rule:
\[
d(uv)=(du)v+q^au(dv),
\]
where $a$ is the grading of the element $u$ and $q$ is a primitive $N-$th
root of unity.

Considering the particular generalization of classical exterior calculus
corresponding to the case $d^3=0$, one replaces the condition $d^2=0$ by $%
d^3=0$ $(d^2\neq 0)$, then one adds to the first order differential of
coordinates $dx^1,dx^2,...dx^n$ a set of second order differentials $%
d^2x^1,d^2x^2,...d^2x^n.$

In the case of a quantum plane such generalization is possible and was
studied in \cite{notre, notre2, bazunova, salih}. The differential algebra
satisfies the covariance property with respect to the quantum group symmetry
of the quantum plane.

In the same spirit, we construct, in the present paper, a $GL_{p,q}(2)$
covariant associative differential algebra, in the two cases $d^2=0$ and $%
d^3=0$ $(d^2 \neq 0)$. We study the application of these differential
calculi to the gauge field theory. In fact, we compute the correspondent
curvatures, from which we extract the gauge field strength.

The paper is organized as follows:

In section $2$ we give the basic definitions and results of the
two-parameter quantum plane. Then we construct an associative differential
calculus with $d^2=0$. We also introduce a generalization of the
differential exterior derivative that satisfies, in stead of the $d^2=0,$
the nilpotency condition $d^3=0$ $(d^2\neq 0)$ in the sense of \cite
{coquereaux1, kerner1}.

In section $3$, the nilpotent endomorphisms introduced in this work, are
used to compute the correspondent curvature expressions on the quantum
plane. Then the field strength which arises from this curvature is deduced.

\section{Differential calculus on the two-parameter quantum plane.}

\subsection{Review of a two-parameter quantum plane.}

The quantum plane \cite{coquereaux1, manin1, manin2} is an unital
associative algebra generated by two non-commuting coordinates $x$ and $y$
satisfying the quadratic relation:

\begin{equation}
xy=qyx, \; \; q\neq 0,1,\;\; q\in C.
\end{equation}

This quantum plane admits a $GL_q(2)$ symmetry, in the sense that the
relation $(1)$ is invariant under its coaction. The main assumption behind
this assertion is the commutation of the coordinates of the quantum plane
and the generators of the quantum group $GL_q(2)$ ( i.e. the entries of the
matrix $T=\left(
\begin{tabular}{ll}
$a$ & $b$ \\
$c$ & $d$
\end{tabular}
\right) $, as a generic element of $GL_q(2)$).

However, if one relaxes this assumption by assuming generic non-commutation
relations between the space coordinates $(x,y)$ and group generators $%
(a,b,c,d)$, the $GL_q(2)$ symmetry is extended to a $GL_{p,q}(2)$ one.

Let us recall \cite{schirrmacher, corrigan} that $GL_{p,q}(2)$ is a quantum
group generated by $a,b,c,d$ obeying the following relations:

\begin{eqnarray}
ab &=&pba\hspace{1.0in}cd=pdc\hspace{1.0in}pbc=qcb  \nonumber \\
ac &=&qca\hspace{1.0in}bd=qdc\hspace{0.68in}ad-da=(p-\frac 1q)bc\,,
\end{eqnarray}
for some non zero complex $p,q$ with $qp\neq -1.$ For latter use and basing
on the results of \cite{notre, notre2} the assumption of non-commutativity
between space coordinates and $GL_{p,q}(2)$ generators takes the form:

\begin{eqnarray}
xa &=&ax\hspace{0.6in} xb =\frac qpbx \hspace{0.7in} xc =cx \hspace{0.6in}
xd =\frac qpdx  \nonumber \\
ya&=&\frac{q}{p}ay \hspace{0.5in}yb=\left( \frac qp\right) ^2by %
\hspace{0.4in} yc=\frac qpcy \hspace{0.5in} yd=\left( \frac qp\right) ^2dy,
\end{eqnarray}
for further details concerning this construction see \cite{schirrmacher,
notre2, corrigan}$.$

In the limit $p\rightarrow q$, we recover the commutativity between
coordinates and generators, and $GL_{p,q}(2)$ reduces to $GL_q(2).$

\subsection{Quantum differential algebra, $d^2=0.$}

The aim of this section is to construct an associative differential algebra $%
^2\Omega _{p,q}=\{x,y,dx,dy\}$, on the two-parameter quantum plane. The
covariance of this differential algebra, with respect to $GL_{p,q}(2)$, is
ensured if we proceed as in \cite{coquereaux1, coquereaux2, notre}$.$

We start by defining the differential operator $d$ satisfying
\[
d(x)=dx,\hspace{1.0in}d(y)=dy,\hspace{1.0in}d(1)=0.
\]
More generally the operator $d$ acts as follows:
\[
d:\Omega ^n\rightarrow \Omega ^{n+1},
\]
where $\Omega ^n$ is the space of forms with degree $n;$ $\Omega ^0$ being
the algebra of functions defined on the quantum plane.

The operator $d$ must also obey the usual properties of:

i/ Linearity:
\[
d(\alpha z+\beta z^{\prime })=\alpha d(z)+\beta d(z^{\prime }),
\]
where $z,$ $z^{\prime }\in $ $^2\Omega _{p,q}$ and $\alpha ,$ $\beta $ are
either complex numbers or the generators $\{a,b,c,d\}$ of $GL_{p,q}(2).$

ii/ Nilpotency:
\[
d^2=0
\]

iii/ Liebniz rule:
\begin{equation}
d(uv)=(du)v+(-1)^nu(dv),
\end{equation}
where $u$ $\in \Omega ^n.$

Moreover, the commutation relations between the elements of $^2\Omega _{p,q}$
must be covariant under $GL_{p,q}(2).$ Namely, one can write a priori $xdx,$
$dy,,$ $ydx$ and $ydy$ in terms of $(dx)x,$ $(dy)x,$ $(dx)y$ and $(dy)y.$
Imposing the covariance of the obtained relations under $GL_{p,q}(2)$,
associativity of expressions such as $(xdx)dy=x(dxdy)$ and differentiating $%
(1)$ yields two possible (mutually related) differential calculi. One of
these calculi was given in \cite{notre2}. The other, which we will use in
the sequel to construct gauge field theory, is given by the following
commutation relations

\begin{eqnarray}
x\,dx &=&qp\;dx\,x~~~~~~~~~~~~~~~~~~x\,dy=q\;dy\,x+(qp-1)dx\,y  \nonumber \\
y\,dy &=&qp\;dy\,y~~~~~~~~~~~~~~~~~~y\,dx=p\;dx\,y \\
dy\,dx &=&-p\;dx\,dy~~~~~~~~~~~~~~(dx)^2=(dy)^2=0,  \nonumber
\end{eqnarray}

The associative differential calculus on the one-parameter quantum plane
\cite{wess1, berezinski, coquereaux1, notre} can be obtained as a limiting
case from the two-parameter one by taking $p\rightarrow q.$

Furthermore, the differential exterior operator $d$ is realized as in the
standard way:

\begin{equation}
d=dx\frac \partial {\partial x}+dy\frac \partial {\partial y}=dx\partial
_x+dy\partial _y,
\end{equation}
where $\partial _x$ and $\partial _y$ are respectively, the partial
derivatives in the directions $x$ and $y.$

Consistency conditions \cite{wess1} then yields:
\begin{eqnarray}
\partial _xx &=&1+qpx\partial _x+(qp-1)y\partial _y\hspace{0.6in}\partial
_yx=qx\partial _y  \nonumber \\
\partial _xy &=&py\partial _x\hspace{1.0in}\hspace{1.0in}\partial
_yy=1+qpy\partial _y \\
\partial _x\partial _y &=&p\partial _y\partial _x\,.  \nonumber
\end{eqnarray}

Latter, we will apply this associative covariant differential calculus to
construct a non commutative gauge field theory on the two-parameter quantum
plane.

The above algebra $^2\Omega _{p,q}$ can be generalized to the one associated
with a nilpotent endomorphism $d$ satisfying $d^3=0.$ This generalization is
defined and studied in the following section.

\subsection{Differential algebra on the two-parameter quantum plane with $%
d^3=0$.}

3-nilpotent differential calculi on specific non commutative spaces (one and
two-parameter quantum plane and superplane) have been constructed in
different papers \cite{notre, notre2, bazunova, salih}.

In a more general context many authors \cite{kerner3, dubois1, coquereaux3}
have studied generalized differential calculi with $d^N=0$ and gave some
examples related to theoretical physics. Here we shall not give an
exposition of these generalizations, but, instead, we shall follow a point
of view which was illustrated in \cite{notre, notre2}$.$ Indeed, one can
always define the differential operator $d$ satisfying linearity, nilpotency
$d^3=0$ $(d^2\neq 0)$ and the $j$-Leibniz rule:
\begin{equation}
d(uv)=(du)v+(j)^nu(dv),\qquad j^3=1,
\end{equation}
where $u\in \Omega ^n$, the space of $n$-forms defined on the two-parameter
quantum plane.

Following the same method adopted in \cite{notre, notre2}, we obtain the
two-parameter associative covariant differential algebra $^3\Omega
_{p,q}=\{x,y,dx,dy,d^2x,d^2y\}.$ However in \cite{notre2} (not in \cite
{notre}) the obtained differential algebra was not associative and we
sketched an adequate way leading to an associative one.

In the present paper, the associativity of the differential algebra is
required in order to construct a gauge field theory. This point will become
clear in the next sections.

Basing on the previous discussion it is easy to show that there will be an
extra constraint on the two parameters $p$ and $q$

\begin{equation}
qp=j^2,
\end{equation}
which arises from the rquirement of the associativity property. And the
commutation relations between the differential generators of $^3\Omega
_{p,q} $ are given by:

First order
\begin{eqnarray}
x \, dx &=& qp \; dx \, x ~~~~~~~~~~~~~~~~~~ x \, dy =q \; dy\, x + (qp-1)
dx \, y  \nonumber \\
y\, dy &=& qp \; dy\, y ~~~~~~~~~~~~~~~~~~ y\, dx =p\; dx\, y
\end{eqnarray}

Second order
\begin{eqnarray}
x\,d^2x &=&qp\;d^2x\,x~~~~~~~~~~~~~~~~~~x\,d^2y=q\;d^2y\,x+(qp-1)d^2x\,y
\nonumber \\
y\,d^2y &=&qp\;d^2y\,y~~~~~~~~~~~~~~~~~~y\,d^2x=p\;d^2x\,y \\
dx\,dy &=&q\;dy\,dx~~~~~~~~~~~~~\;\,~~~~dy\,dx=jp\;dx\,dy  \nonumber
\end{eqnarray}

Third order
\begin{eqnarray}
dx\,d^2x&=&j\; d^2x\,dx ~~~~~~~~~~~~~~~~~~ dx\,d^2y=j^2q\;
d^2y\,dx+j^2(qp-1)\;d^2x\,dy  \nonumber \\
dy\,d^2y&=&j\; d^2y\,dy ~~~~~~~~~~~~~~~~~~ dy\,d^2x=j^2p\; d^2x\,dy
\end{eqnarray}

Fourth order
\begin{equation}
d^2x\,d^2y=q\;d^2yd^2x.
\end{equation}

\vspace{0.5cm}

Using the realization of the differential operator $d:$
\[
d=dx\partial _x+dy\partial _y,
\]
one can prove that:

\begin{eqnarray}
\partial _xx&=&1+qpx\partial _x+(qp-1)y\partial _y\hspace{0.6in}\partial
_yx=qx\partial _y  \nonumber \\
\partial _xy&=& py\partial _x\hspace{1.0in}\hspace{1.0in}\partial
_yy=1+qpy\partial _y \\
\partial _x\partial _y&=&p\partial _y\partial _x \hspace{1.8in}%
(dx)^3=(dy)^3=0.  \nonumber
\end{eqnarray}

The natural requirement to recover the one-parameter differential calculus
when $p\rightarrow q$ is preserved. In fact, in this limit one has $q=j$,
due to $(9)$, and all the results reduce to their counterparts obtained in
\cite{notre}; especially, the two-parameter quantum plane becomes a reduced
quantum plane.

As a physical application of these differential calculi `$d^2=0$' and `$%
d^3=0 $' we construct, in the sections below, the corresponding gauge field
theory.

\section{Covariant derivative and curvature on a two-parameter quantum plane.
}

\subsection{`$d^2=0$' case.}

Here, our main purpose is to define the covariant derivative; as in the
ordinary case, this enables us to introduce the notion of curvature, the
components of which, will be discussed by comparing them to the ordinary
ones.

So, as in the commutative case, the covariant differential acting on a field
$\Phi (x,y)$ is defined by:

\begin{equation}
D\Phi (x,y)=d\Phi (x,y)+A(x,y)\Phi (x,y),
\end{equation}
where $\Phi (x,y)$ is a function on the two-parameter quantum plane and the
gauge field $A(x,y)$ is a 1-form valued in the associative algebra of
functions on the two-parameter quantum plane.

We notice that the bimodule structure of the algebra of functions on the
two-parameter quantum plane over $^2\Omega _{p,q}$ is assumed.

As usual, the covariant differential $D$ must satisfy

\begin{equation}
DU^{-1}\Phi (x,y)=U^{-1}D\Phi (x,y),
\end{equation}
where $U$ is an endomorphism defined on the algebra of functions over the
two-parameter quantum plane.

From $(16)$ it follows that the gauge field transforms as:

\begin{equation}
A(x,y)\rightarrow U^{-1}A(x,y)U+U^{-1}dU.
\end{equation}

The curvature is defined through:

\begin{equation}
D^2\Phi (x,y)=\left( dA(x,y)+A(x,y)A(x,y)\right) \Phi (x,y):=R\Phi (x,y).
\end{equation}

The gauge field $A(x,y)$ being a 1-form, generally takes the form:

\begin{equation}
A(x,y)=dx\,\,A_x(x,y)+dy\,\,A_y(x,y),
\end{equation}
where the component $A_x(x,y)$ and $A_y(x,y)$ are functions on the
two-parameter quantum plane.

$Eq$(19), together with the differential realization of $d$ $(6)$ and taking
account of the associativity of $^2\Omega _{p,q}$, allows to rewrite the
curvature $R$:

\begin{eqnarray}
R &=&dxdy\,\left( \frac 1p\partial _xA_y(x,y)-\partial _yA_x(x,y)\right)
+dxA_x(x,y)dxA_x(x,y)+  \nonumber \\
&&+dxA_x(x,y)dyA_y(x,y)+dyA_y(x,y)dxA_x(x,y)+dyA_y(x,y)dyA_y(x,y).
\end{eqnarray}

One has to express the curvature written above in terms of 2-forms
constructed from the basic generators of the differential algebra $^2\Omega
_{p,q}$. Since we are dealing with a non-commutative space (two-parameter
quantum plane), this task is not straightforward. In fact, the
non-commutativity prevents us from rearranging the different terms in $(20)$
adequately. To overcome this technical difficulty we require that the
components of the gauge field, $A_x(x,y)$ and $A_y(x,y)$, are expressed as
formal power series of the space coordinates:

\begin{eqnarray}
A_x(x,y) &=&a_{n,m} y^nx^m \\
A_y(x,y) &=&b_{n,m}y^nx^m,  \nonumber
\end{eqnarray}
where $n$ and $m$ are integers. Using the formulas $(1),(5),(7),(20),(21)$
and after technical computations, an explicit expression of curvature arises:

\begin{equation}
R=dxdy\ \left[ F_{xy}^{\frac 1p}-\frac 1p\ \ \alpha
^{n,m}b_{n,m}y^{n+1}x^{m-1}A_y(x,y)\right] ,
\end{equation}
where
\begin{eqnarray}
F_{xy}^{\frac 1p}=\frac 1p\partial _xA_y(x,y)-\partial
_yA_x(x,y)+A_x(qx,qpy)A_y(x,y)-\frac 1pA_y(qpx,py)A_x(x,y)
\end{eqnarray}
and
\begin{equation}
\alpha ^{n,m}=p^nq^{m-1}\ \left( (qp)^m-1\right) .
\end{equation}

It is easy to check that the deformed field strength $F_{xy}^{\frac 1p}$
satisfy a $p$-antisymmetry property:
\[
F_{yx}^p = -p \; F_{xy}^{\frac 1p} \; .
\]

If one takes the limit $p\rightarrow q=j$, the curvature expression obtained
in \cite{notre} is recovered. More interesting is the limit $p\rightarrow
q\rightarrow 1$, where one expects to recover the commutative case. This is
valid:

\[
F_{xy}^{\frac 1p}\rightarrow F_{xy}\ \ \hbox{since\quad }\alpha
^{n,m}\rightarrow 0.
\]
This remark allows the interpretation of the supplementary terms $\left(
-\frac 1p\ \alpha ^{n,m}b_{n,m}y^{n+1}x^{m-1}A_y(x,y)\right) $, appearing in
$(22)$, as a direct consequence of the non-commutativity of the space.

\subsection{`$d^3=0$' case.}

In this section we extend the results of the previous one in the sense that
we will use the differential calculus `$d^3=0$'.

The same definition $(15)$ of the covariant differential as well as
properties $(16),(17),(19)$ are preserved in this case \cite{kerner3}.

However, the curvature is no more a 2-form defined in $(18)$ but is a 3-form
\cite{kerner2, kerner3}:

\begin{equation}
D^3\Phi (x,y):=R\Phi (x,y).
\end{equation}
Direct computations show that $R$ is given by:

\begin{eqnarray}
R &=&d^2A(x,y)+dA^2(x,y)+A(x,y)dA(x,y)+A^3(x,y)  \nonumber \\
\ &=&d^2A(x,y)+dA(x,y)A(x,y)-j^2A(x,y)dA(x,y)+A^3(x,y).
\end{eqnarray}

As in the previous case $(d^2=0)$, we will use the formal power series
expansion $(21)$ of the gauge field. Then general formulas (given in
appendix A) can be obtained using $(1),(10)-(14).$ These formulas simplify
the computation of the curvature $R$ and permit to write it in terms of the
basic generators of $^3\Omega _{p,q}:$

\begin{eqnarray}
R
&=&jd^2xdyF_{xy}^p+d^2ydxF_{yx}^{jq}+(dx)^2dy(R_{xxy}+R_{xyx}+R_{yxx})+(dy)^2dx(R_{yyx}+R_{yxy}+R_{xyy})
\nonumber \\
&&\quad -jpd^2xdyA_0^b(x,y)A_y(x,y)+d^2ydxA_0^b(x,y)A_y(x,y)  \nonumber \\
&&\quad +(dx)^2dy\Big[\left\{
j^2pA_2(x,y)+p^2A_1^b(x,y)+(jp)^2A_0^b(qpx,py)A_0^b(x,y)\right\} A_y(x,y)
\nonumber \\
&&\quad
+jpA_x(q^2px,qp^2y)A_0^b(x,y)A_y(x,y)+q(jp)^2A_0^b(qpx,py)A_x(qx,qpy)A_y(x,y)
\nonumber \\
&&\quad +(jp)^2A_0^b(qpx,py)A_y(qpx,py)A_x(x,y)  \nonumber \\
&&\quad +\left\{ (1+jp^2)A_0(qx,qpy)+(jp)^2A_3^b(x,y)\right\}
A_y(qx,qpy)A_y(x,y)  \nonumber \\
&&\quad +(jp)^2A_y((qp)^2x,p^2y)A_0(x,y)A_y(x,y)  \nonumber \\
&&\quad -\left\{ (1+jp^2)A_0(qx,qpy)+(jp)^2A_3^b(x,y)\right\} \partial
_yA_y(x,y)  \nonumber \\
&&\quad -A_0^b(qpx,py)\left\{ q(jp)^2\partial _yA_x(x,y)+qp^3\partial
_xA_y(x,y)\right\} \Big]  \nonumber \\
&&\quad +(dy)^2dx\Big[%
jA_2^b(x,y)A_y(x,y)+(q+p)A_0^b(qx,qpy)A_y(qx,qpy)A_y(x,y)  \nonumber \\
&&\quad +A_y(q^2px,qp^2y)A_0^b(x,y)A_y(x,y)-(q+p)A_0^b(qx,qpy)\partial
_yA_y(x,y)\Big],
\end{eqnarray}
where the field strength is given by
\begin{eqnarray}
F_{xy}^p &=&p\partial _xA_y(x,y)-\partial
_yA_x(x,y)+A_x(qx,qpy)A_y(x,y)-pA_y(qpx,py)A_x(x,y)  \nonumber \\
F_{yx}^{jq} &=&jq\partial _yA_x(x,y)-\partial
_xA_y(x,y)+A_y(qpx,py)A_x(x,y)-jqA_x(qx,qpy)A_y(x,y).
\end{eqnarray}
The components $R_{ijk},i,j,k=x,y$, and the functions $%
A_0,A_0^b,A_1^b,A_2^b,A_3^b$ are given in appendix B.

The antisymmetric property of the field strength $F_{xy}$ in the commutative
case, is replaced by a $p$-antisymmetry:

\begin{equation}
F_{xy}^p=-pF_{yx}^{jq}
\end{equation}

In the limit $p\rightarrow q$ the components $F_{xy}^p,F_{yx}^{jq},R_{ijk}$
and $A_0,A_0^b,A_1^b,A_2^b,A_3^b$ reduce to their one-parameter counterparts
obtained in \cite{notre}; then the curvature will be identical to the one in
\cite{notre}.

The expression of the curvature components $(31)$ in appendix B, and the
deformed field strength $(28)$ are formally the same as those obtained by
\cite{kerner2, kerner3}. The supplementary terms, where the functions $%
A_0,A_0^b,A_1^b,A_2^b,A_3^b$ appear, can be interpreted as a direct
consequence of the non-commutativity property of the space.

Moreover, compared with the case $d^2=0$, the curvature expression contains
additional terms $R_{ijk}$. These terms can be interpreted as a generic
consequence of the generalization of the differential calculus $d^2=0$ to a
higher order $d^3=0$.

\section{Conclusion}

In this work we have constructed associative differential calculi $d^2=0, \;
d^3=0$ on the two-parameter quantum plane. The notion of covariance of these
differential calculi is also ensured and we have shown that there is a
quantum group, $GL_{p,q}(2),$ behind this covariance.

As an application, we have constructed a gauge field theory based on these
differential calculi. The limit $p\rightarrow q$ was also studied in the two
cases $d^2=0,\; d^3=0$ and yields the results of \cite{notre}.

\section*{Appendix A}

In this appendix we give some general formulas which are useful in the
computation of the curvature components.
\begin{eqnarray}
x^n\, dx \!\!\! &=& \!\!\!(qp)^n \; dx \, x^n
~~~~~~~~~~~~~~~~~~~~~~~~~~~~~~~ x^n \, dy =q^n \; dy \, x^n + q^{n-1}\Big(%
(qp)^n-1\Big)\; dx \, y\, x^{n-1}  \nonumber \\
&&  \nonumber \\
y^n \, dx \!\!\! &=& \!\!\! p^n \; dx \, y^n
~~~~~~~~~~~~~~~~~~~~~~~~~~~~~~~~~~~ y^n \, dy = (qp)^n \; dy \, y^n
\nonumber \\
&&  \nonumber \\
\partial _x(y^nx^m)\!\!\! &=& \!\!\! p^n \; \left(\frac{1-(qp)^m}{1-qp}%
\right) \; y^n \, x^{m-1}~~~~~~~ \partial _y(y^nx^m)= \left( \frac{1-(qp)^n}{%
1-qp}\right) \; y^{n-1}\, x^m \\
&&  \nonumber \\
A_z(x,y)\, dx \!\!\! &=& \!\!\! dx\, A_z(qpx,py) ~~~~~~~~~~~~~~~~~~~
A_z(x,y)\, dy = dy\, A_z(qx,qpy)+ dx\; \alpha^{n,m} c_{n,m}\; y^{n+1} \,
x^{m-1}  \nonumber \\
&&  \nonumber \\
\partial _xA_z(x,y) \, dx \!\!\! &=& \!\!\! dx\, \partial _xA_z\mid
_{(qpx,py)}~~~~~~~~~~~~~~ \partial _xA_z(x,y)\, dy = dy \, \partial
_xA_z\mid _{(qx,qpy)} + dx \; \beta_x^{n,m} c_{n,m}\; y^{n+1} \, x^{m-2}
\nonumber \\
&&  \nonumber \\
\partial _yA_z(x,y) \, dx \!\!\! &=& \!\!\! dx \; \partial
_yA_z\mid_{(qpx,py)} ~~~~~~~~~~~~~~ \partial _yA_z(x,y)\, dy = dy \,
\partial _yA_z\mid_{(qx,qpy)} + dx \; \beta_y^{n,m}c_{n,m} \; y^n\, x^{m-1},
\nonumber
\end{eqnarray}
where $z=x,y$ and $c_{n,m}=a_{n,m}$ for $z=x$ or $c_{n,m}=b_{n,m}$ for $z=y$.

\section*{Appendix B}

In this appendix we give the explicit expression of the curvature components
appearing in $(24)$

$R_{xxy}=(jp)^2\; \partial _x\partial _xA_y(x,y)+j\; \partial _xA_x\left|
_{(qx,qpy)}\right. A_y(x,y)- jp \; A_x(q^2px,qp^2y)\, \partial _x \,
A_y(x,y)+$
\[
A_x(q^2px,qp^2y) \, A_x(qx,qpy) \, A_y(x,y)
\]

$R_{yxx}=\partial _y\partial _xA_x(x,y)+j^2p \; \partial _yA_x\left|
_{(qpx,py)}\right. A_x(x,y)-(jp)^2 \; A_y((qp)^2x,p^2y)\, \partial _x \,
A_x(x,y)+$
\[
(jp)^2 \; A_y((qp)^2x,p^2y) \, A_x(qpx,py) \, A_x(x,y)
\]

$R_{xyx}=jp \; \partial _x\partial _yA_x(x,y)+p^2 \; \partial _xA_y\left|
_{(qpx,py)}\right. A_x(x,y) - A_x(q^2px,qp^2y)\, \partial _y \, A_x(x,y)+$
\begin{equation}
jp\; A_x(q^2px,qp^2y) \, A_y(qpx,py) \, A_x(x,y)
\end{equation}

$R_{yyx}=q^2 \; \partial _y\partial _yA_x(x,y)+j \; \partial _yA_y\left|
_{(qpx,py)}\right. A_x(x,y)- q \; A_y(q^2px,qp^2y)\, \partial _yA_x(x,y)+$
\[
A_y(q^2px,qp^2y)\, A_y(qpx,py) \, A_x(x,y)
\]

$R_{yxy}=q \; \partial _y\partial _xA_y(x,y)+jq^2 \; \partial _yA_x\left|
_{(qx,qpy)}\right. A_y(x,y)-jqp \; A_y(q^2px,qp^2y)\, \partial _xA_y(x,y)+$
\[
q\; A_y(q^2px,qp^2y)\, A_x(qx,qpy) \, A_y(x,y)
\]

$R_{xyy}=\partial _x\partial _yA_y(x,y)+j^2q^2p \; \partial _xA_y\left|
_{(qx,qpy)}\right. A_y(x,y)-q^2\; A_x(q^2x,(qp)^2y) \, \partial _yA_y(x,y)+$
\[
q^2\; A_x(q^2x,(qp)^2y) \, A_y(qx,qpy) \, A_y(x,y) \, .
\]

The functions $A_0,A_0^b,A_1^b,A_2^b,A_3^b$ are given by
\begin{eqnarray}
A_0 \!\!\! &=& \!\!\! \alpha^{n,m}a_{n,m} \; y^{n+1}\, x^{m-1}
~~~~~~~~~~~~~~~~ A_0^b = \alpha ^{n,m}b_{n,m} \; y^{n+1} \, x^{m-1}
\nonumber \\
A_1^b \!\!\! &=& \!\!\! \beta_x^{n,m}b_{n,m} \; y^{n+1} \, x^{m-2}
~~~~~~~~~~~~~~~~\, A_2^b = \beta_y^{n,m} b_{n,m} \; y^n \, x^{m-1} \\
A_3^b \!\!\! &=& \!\!\! \alpha^{n,m} b_{n,m} q^{m-2} p^{n+1}\Big((qp)^{m-1}-1%
\Big) \; y^{n+2}\, x^{m-2},  \nonumber
\end{eqnarray}

where the coefficients $\alpha^{n,m},$ $\beta_x^{n,m}$ and $\beta _y^{n,m}$
are given by:
\begin{eqnarray}
\alpha^{n,m} &=& q^{m-1}p^n \; \Big((qp)^m-1 \Big)  \nonumber \\
\beta_x^{n,m} &=& q^{m-2}p^{2n} \Big((qp)^{m-1}-1\Big) \left( \frac{1-(qp)m}{%
1-qp}\right) \\
\beta_y^{n,m} & =& q^{m-1}p^{n-1}\Big((qp)^m-1\Big)\left(\frac{1-(qp)^n}{1-qp%
}\right) \;  \nonumber
\end{eqnarray}

\end{document}